\documentclass{jdsart2}
 
 \volume{0}
 \issue{0}
 \pubmonth{July}
 \pubyear{2024}
 \articletype{research-article}
 \doi{\empty}
 \firstpage{1}
 \lastpage{11}
 
 \theoremstyle{plain}

 \theoremstyle{remark}
 
 \theoremstyle{definition}
 
\usepackage{bm, amsmath,latexsym,amsthm,amssymb,amsbsy,amsfonts,lscape,ctable, mathtools}
\usepackage{algorithm, algorithmic}
\usepackage{natbib}
\usepackage{subfig}
\usepackage[english]{babel}
\newtheorem{theorem}{Theorem}
\DeclareMathOperator\erf{erf}

\DeclareMathOperator*{\argmin}{argmin} % thin space, limits underneath in displays
\renewcommand\vec{\mathbf}
\usepackage{listings} % Package for code listings
\usepackage{color}
\usepackage{fancyvrb}

\DefineVerbatimEnvironment{Highlighting}{Verbatim}{commandchars=\\\{\}}
\usepackage{framed}
\definecolor{shadecolor}{RGB}{248,248,248}
\newenvironment{Shaded}{\begin{snugshade}}{\end{snugshade}}

\newcommand{\AttributeTok}[1]{\textcolor[rgb]{0.13,0.29,0.53}{#1}}

\newcommand{\CommentTok}[1]{\textcolor[rgb]{0.56,0.35,0.01}{\textit{#1}}}

\newcommand{\ConstantTok}[1]{\textcolor[rgb]{0.56,0.35,0.01}{#1}}
\newcommand{\ControlFlowTok}[1]{\textcolor[rgb]{0.13,0.29,0.53}{\textbf{#1}}}

\newcommand{\DecValTok}[1]{\textcolor[rgb]{0.00,0.00,0.81}{#1}}

\newcommand{\FloatTok}[1]{\textcolor[rgb]{0.00,0.00,0.81}{#1}}
\newcommand{\FunctionTok}[1]{\textcolor[rgb]{0.13,0.29,0.53}{\textbf{#1}}}

\newcommand{\NormalTok}[1]{#1}

\newcommand{\OtherTok}[1]{\textcolor[rgb]{0.56,0.35,0.01}{#1}}

\newcommand{\SpecialCharTok}[1]{\textcolor[rgb]{0.81,0.36,0.00}{\textbf{#1}}}

\newcommand{\StringTok}[1]{\textcolor[rgb]{0.31,0.60,0.02}{#1}}

\begin{document}
\begin{frontmatter}
\pretitle{Research Article}
 
\title{Unified robust boosting}
\runtitle{Unified robust boosting}
 
\author[1]{\inits{Z.}\fnms{Zhu} \snm{Wang}\thanksref{c1}\ead{zwang145@uthsc.edu}}
\thankstext[type=corresp,id=c1]{}
%\thankstext[type=corresp,id=c1]{Corresponding author.}
% \thankstext[id=f1]{Simple footnote.}
 \address[1]{Memphis, TN,
 \institution{The University of Tennessee Health Science Center}, \cny{United States}}
 \begin{abstract}
Boosting is a popular algorithm in supervised machine learning with wide applications in regression and classification problems. It combines weak learners, such as regression trees, to obtain accurate predictions. However, in the presence of outliers, traditional boosting may yield inferior results since the algorithm optimizes a convex loss function. Recent literature has proposed boosting algorithms that optimize robust nonconvex loss functions. Nevertheless, there is a lack of weighted estimation to indicate the outlier status of observations.
This article introduces the iteratively reweighted boosting (IRBoost) algorithm, which combines robust loss optimization and weighted estimation. It can be conveniently constructed with existing software. The output includes weights as valuable diagnostics for the outlier status of observations. For practitioners interested in the boosting algorithm, the new method can be interpreted as a way to tune robust observation weights.
IRBoost is implemented in the \proglang{R} package \pkg{irboost} and is demonstrated using publicly available data in generalized linear models, classification, and survival data analysis.
\end{abstract}
 
\end{frontmatter}

\section{Introduction}
Boosting is a powerful supervised machine learning algorithm. As an ensemble method, boosting combines many weak learners to generate a strong prediction. Being a functional descent method, boosting finds wide applications in regression and classification problems. \citet{Frie:2001} and \citet{friedman2000additive} discussed boosting for a variety of convex loss functions. Boosting can be utilized to fit various models with different base learners, including linear least squares, smoothing splines, and regression trees \citep{Buhl:Hoth:2007, wang2018robust}.

The \proglang{R} package \pkg{mboost} implements robust boosting with the Huber loss for continuous responses \citep{hothorn2023mboost}. Likewise, the package \pkg{xgboost} \citep{chen2024xgboost} implements a pseudo-Huber loss function. Notably, these loss functions are convex. While convex loss functions have computational advantages and are commonly used due to their ease of optimization, they can lack robustness to leverage points in the predictor variables \citep{maronna2019robust} and provide poor approximations to certain loss functions like the 0-1 loss \citep{wu2007robust, zhao2010convex, park2011robust} in classification problems.

On the other hand, nonconvex loss functions, such as Tukey's biweight loss, can be robust to both vertical outliers in the response variable and leverage points. Moreover, nonconvex loss functions can achieve better generalization accuracy.

Recent research has explored strategies to address outliers in boosting algorithms with nonconvex loss functions. For example, a study on boosting in the presence of outliers examined the role and efficiency of nonconvex loss functions for binary classification problems, aiming to adaptively handle outliers during the boosting process \citep{li2018boosting}. \citet{wang2018quadratic} and \citet{wang2018robust} proposed robust functional gradient boosting for nonconvex loss functions in the
context of regression and classification problems, implemented in the \proglang{R} package \pkg{bst} \citep{wang2023bst}. These methods applied a majorization-minimization (MM) scheme, an extension of the popular expectation-maximization (EM) algorithm in statistics. These MM algorithms involve quadratic
 approximations of functions and difference-of-convex functions. However, they fail to generate weights as an indication of the outlier status of the observations. Small weights should be assigned to observations deviating from the underlying model.

In classical robust estimation, weights are derived from robust loss functions, such as the Huber loss. Recent progress has been made in generating weights from robust loss functions in more complex problems. \citet{wang2024unified} innovatively proposed a new framework for robust estimation by reducing the weight of the observation that leads to a large loss. The author introduced a unified class of robust loss functions known as the concave-convex (CC) family. Additionally, the author proposed iteratively reweighted convex optimization (IRCO), a specialized application of the MM algorithm designed to minimize the loss functions within the CC family. The CC-family includes traditional robust loss functions such as the Huber loss, robust hinge loss for support vector machines, and robust exponential family for generalized linear models. IRCO can be conveniently implemented with existing methods and software.

In this article, we integrate IRCO and boosting into IRBoost for the CC-family. This functional optimization is more general than the parameter-based estimation in \citet{wang2024unified}. For instance, IRBoost permits a function space derived from regression trees. Unlike previous robust boosting methods, including \citet{wang2018quadratic, wang2018robust}, the major novelty is that IRBoost provides a unified framework for a large class of robust loss functions to estimate weights and identify outliers.

We illustrate the proposed algorithm using the \proglang{R} package \pkg{irboost} \citep{wang2024irboost}, applying it to various models within the robust exponential family, such as regression, logistic regression, and Poisson regression. Additionally, we demonstrate its application to robust survival regression with the accelerated failure time model. The package further includes implementations of IRBoost for gamma regression, Tweedie regression, hinge classification, and multinomial logistic regression.
\section{Robust boosting}\label{sec:robust}
\subsection{CC-family function estimation}\label{sec:dcb}
To unify robust estimation, \citet{wang2024unified} proposed the concave convex family with functions $\Gamma=g\circ s$ satisfying the following conditions:
\begin{enumerate}
\item $g$ is a nondecreasing closed concave function whose domain is the range of function $s$.
\item $s$ is convex on $\mathbb{R}$.
\end{enumerate}
Examples of concave component $g$ are listed in Table~\ref{tab:gs}. 
Note that the function \code{tcave} is not differentiable everywhere but subdifferentiable. The parameter $\theta$ controls robustness level a model is allowed to have, and a smaller value leads to a more robust estimation. 
See \citet{wang2024unified} for a discussion of the motivation behind these functions. In classification problems, we assume $y \in \{-1, 1\}$.
\begin{table}
\begin{center}
\renewcommand{\arraystretch}{1} % Default value: 1
\begin{tabular}{lc}
\hline\hline
Concave function& $g(z), z \geq 0$\\
\hline
\code{hcave}&$\begin{cases}
z & \text { if } z \leq \theta^2/2\\ 
\theta (2z)^{\frac{1}{2}}-\frac{\theta^2}{2} &\text{ if } z > \theta^2/2
\end{cases}$ \\
    \code{acave}&\hspace{3mm}$\begin{cases}
{\theta^2}(1-\cos(\frac{(2z)^{\frac{1}{2}}}{{\theta}})) & \text{ if } z \leq \theta^2\pi^2/2\\
2\theta^2 &\text{ if } z > \theta^2\pi^2/2\\
\end{cases}$
\\
        \code{bcave}&\hspace{3mm}$\frac{\theta^2}{6}\left(1-(1-\frac{2z}{\theta^2})^3 I(z \leq \theta^2/2)\right)$  \\
        \code{ccave}&\hspace{3mm}$\theta^2\left(1-\exp(\frac{-z}{\theta^2})\right)$ \\
        \code{dcave}&\hspace{3mm}$\frac{1}{1-\exp(-\theta)}\log(\frac{1+z}{1+z\exp(-\theta)})$ \\
        \code{ecave}&\hspace{3mm}$\begin{cases}
\frac{2\exp(-\frac{\delta}{\theta})}{\sqrt{\pi\theta\delta}}z &\text{ if }z\leq \delta\\
\erf(\sqrt{\frac{z}{\theta}})-\erf(\sqrt{\frac{\delta}{\theta}})+\frac{2\exp(-\frac{\delta}{\theta})}{\sqrt{\pi\theta\delta}}\delta &\text{ if } z>\delta
\end{cases}$\\
            \code{gcave}&\hspace{3mm}
$\begin{cases}
\frac{\delta^{\theta-1}}{(1+\delta)^{\theta+1}}z &\text{ if } z \leq \delta\\
\frac{1}{\theta}(\frac{z}{1+z})^{\theta}-\frac{1}{\theta}(\frac{\delta}{1+\delta})^{\theta}+\frac{\delta^{\theta}}{(1+\delta)^{\theta+1}} &\text{ if }z   > \delta
\end{cases}$ \\
&where
$\delta=
\begin{cases}
\to 0+ &\text{ if } 0 < \theta < 1\\ 
\frac{\theta-1}{2} &\text{ if }\theta \geq 1
\end{cases}
$\\
\code{tcave}&\hspace{3mm}$\min(\theta, z), \theta \geq 1 \text{ for classification; }\theta > 0 \text{        otherwise }$ \\
\hline
\hline
\end{tabular}
\end{center}
\caption{Concave component with $\theta > 0$.} 
\label{tab:gs}
\end{table}
The convex component includes common loss functions in regression and classification such as squared loss $s(u)=u^2$ and the negative log-likelihood function in the exponential family adopted by generalized linear models. Other examples include negative log-likelihood functions for multinomial logistic regression, Tweedie regression, and the accelerated failure time model for time-to-event data subject to censoring \citep{barnwal2022survival}.

Given a set of observations $(\vec x_i, y_i), i=1, ..., n$, where $y_i \in \mathbb{R}$ and $\vec x_i = (x_{i1}, ..., x_{ip})^\mathsf{T} \in \mathbb{R}^p$, denote $\Omega$ as the linear span of a set $H$ of base learners, including regression trees and linear predictor functions. An estimation of $\vec f=(f(\vec x_1), ..., f(\vec x_n))^\mathsf{T} \in \Omega$ can be obtained by minimizing an empirical loss function
\begin{equation}\label{eqn:emlos}
\sum_{i=1}^n \ell\biggl(y_i, f(\vec x_i)\biggr),
\end{equation}
where $\ell$ is a CC-family member, $\ell=g\circ s=g(s(u))$. With some abuse of notation, $s(u)$ is also used to denote $s(y, f(\vec x))$. For instance, $s(u)=s(y-f(\vec x))$ in regression, and 
$s(u)=s(yf(\vec x))$ in a margin-based classification. 
To simplify the notation, $f$ is often used to replace $f(\vec x)$. 

The robust function estimation problem (\ref{eqn:emlos}) can be addressed using Algorithm~\ref{alg:ccf}, employing an iteratively reweighted boosting approach. This two-layer algorithm operates as follows: in each iteration at the outer layer, observation weights are updated based on the current loss values from the convex component. With these weights, the inner layer updates function estimation and loss values from the convex component through weighted boosting iterations minimizing
weighted convex loss functions. In a new cycle, the outer layer weights are updated using the boosting results. Subsequently, the inner layer runs new boosting iterations with the updated weights. The process repeats until convergence.
This approach represents a generalization of the iteratively reweighted least squares method commonly employed in robust estimation \citep{maronna2019robust, wang2024unified}.

Step 3 involves $\varphi$, the Fenchel conjugate of $-g$ defined as:
\begin{equation*}
    \varphi(v) = \sup_{z \in \text{dom } g}\biggl(zv+g(z)\biggr).
\end{equation*}
Here, $\partial (-g(z))$ means the subdifferential of function $-g$ at point $z$, which is  a set of slopes that touch the graph of $-g$ at $(z, -g(z))$ and bound the graph from below. If the function is differentiable at $z$, then $\partial (-g(z))$ contains only one element ${-g'(z)}$. Let 
\begin{equation*}
\rho(\vec f^{(k)})=\sum_{i=1}^n\ell(y_i, f_i^{(k)}),
\end{equation*}
where $\vec f^{(k)}$ is generated by the algorithm.
We then have the following convergence results for IRBoost.
\begin{algorithm}[!htbp]
\begin{algorithmic}[1]
\caption{IRBoost}\label{alg:ccf}
\STATE \textbf{Input:} training samples $\{(\vec x_1, y_1), ..., (\vec x_n, y_n)\}$, 
concave component $g$ with parameter $\theta$, convex component $s$, starting points $z_i, i=1, ..., n$ and iteration count $K$. 
% \STATE Set $\delta > \epsilon$ for a pre-specified small value $\epsilon > 0 $ for convergence criteria.
\FOR{$k=1$ to $K$}
%    \STATE \textbf{Initialization:} ${f}_0(x)=f^{(k-1)}$. 
    \STATE Compute subgradient $v_i^{(k)}$ via $v_i^{(k)}\in \partial(-g(z_i))$ or $z_i \in \partial \varphi(v_i^{(k)}), i=1, ..., n$.
\STATE Compute $\vec f^{(k)}=\argmin_{\vec f \in \Omega} \sum_{i=1}^n s(y_i, f_i)(-v_i^{(k)})$.
\STATE %Compute $u_i^{(k)}$ in (\ref{eqn:ui}) based on the current $f_i^{(k-1)}$, 
Compute $z_i=s(y_i, f_i^{(k)}), i=1, ..., n$.
\ENDFOR
\STATE \textbf{Output:} $v_i^{(K)}$ and $\vec f^{(K)}$. 
\end{algorithmic}
\end{algorithm}
\begin{theorem}\label{thm:conv1}
Suppose that $g$ is a concave component in the CC-family, and $g$ is bounded below.
%$z_i$ is an interior point of $\textup{dom } g$ or $v_i^{(k+1)}$ is an interior point of $\textup{dom } \varphi$.
Then the loss function values $\rho(\vec f^{(k)})$ generated by Algorithm~\ref{alg:ccf} are nonincreasing and converge.
\end{theorem}

This result extends Theorem 6 in \citet{wang2024unified}, which focused on linear predictor functions. Our investigation encompasses more expansive function spaces. Conversely, when $\Omega$ represents a linear space of linear models, Theorem~\ref{thm:conv1} aligns with the results detailed in \citet{wang2024unified}. Algorithm~\ref{alg:ccf} is an MM algorithm, and the proof follows the same argument as Theorem 6 in \citet{wang2024unified}. Hence, only a sketch of the proof is given below, specifically assuming that $g$ is differentiable.

For a differentiable concave function $g$, the first-order condition is $\forall u, v \in \text{dom } g$
\begin{equation}\label{eqn:foc}
    g(u) \leq g(v) + g'(v)(u-v).
\end{equation}
Substitute $u$ with $s(u)$, and $v$ with $s(v)$ in (\ref{eqn:foc}), we get
\begin{equation}\label{eqn:foc1}
    g(s(u)) \leq g(s(v)) + g'\left(s(v))(s(u)-s(v)\right).
\end{equation}
Substitute $s(u)=s(y_i, f_i), s(v)=s(y_i, f_i^{(k)})$ in (\ref{eqn:foc1}), and sum up for $i=1, ..., n$, we get 
\begin{equation}\label{eqn:foc2}
    \sum_{i=1}^n g(s(y_i, f_i)) \leq \sum_{i=1}^n g\left(s(y_i, f_i^{(k)})\right)+g'\left(s(y_i, f_i^{(k)})\right)\left(s(y_i, f_i)-s(y_i, f_i^{(k)})\right).
\end{equation}
Let $Q(\vec f|\vec f^{(k)})$ denote the right hand side of (\ref{eqn:foc2}). We then have
\begin{equation}\label{eqn:mm1}
\rho(\vec f) \leq Q(\vec f|\vec f^{(k)}), \quad \rho(\vec f^{(k)})=Q(\vec{f}^{(k)}|\vec{f}^{(k)}).
\end{equation}
This implies that $Q(\vec f|\vec f^{(k)})$ majorizes $\rho(\vec f)$ at $\vec f^{(k)}$. The algorithm iterates as follows: given an estimate $\vec f^{(k)}$ in the $k$th iteration, $Q(\vec f|\vec f^{(k)})$ is minimized in the $k+1$ iteration to obtain an updated minimizer $\vec f^{(k+1)}$. This process is repeated until convergence. Therefore, the algorithm generates a descent sequence of estimates:
\begin{equation}\label{eqn:mm6}
\rho(\vec f^{(k+1)}) \leq Q(\vec f^{(k+1)}|\vec f^{(k)}) \leq            Q(\vec{f}^{(k)}|\vec{f}^{(k)}) 
=\rho(\vec f^{(k)}). 
\end{equation}
Alternatively, the majorization (\ref{eqn:mm1}) can be constructed 
from a different surrogate function derived from the Fenchel convex conjugate. 
The Fenchel-Moreau theorem states that the following result holds \citep{wang2024unified}
\begin{equation*}
    g(s(u))=\inf_{v \in \text{dom }\varphi}\left(s(u)(-v)+\varphi(v)\right).
\end{equation*}
Let
\begin{equation*}
R(\vec f|\vec f^{(k)})=\sum_{i=1}^ns(y_i, f_i)(-v_i^{(k)})+\varphi(v_i^{(k+)}). 
\end{equation*}
$R(\vec f|\vec f^{(k)})$ is another function that majorizes $\rho(\vec f)$ at $\vec f^{(k)}$. The algorithm generates a sequence of estimates in a descending order, similar to (\ref{eqn:mm6}).

In step 3 of IRBoost, weights are computed using two different methods, each associated with a surrogate function, namely $Q$ and $R$. Remarkably, the solutions obtained from these different approaches have been demonstrated to be identical \citep{wang2024unified}.
%in the algorithm is equivalent to minimizing $Q(\vec f|\vec f^{(k)})$ since $\varphi(v_i^{(k+1)})$ is a constant with respect to $\vec f$. 

\subsection{Boosting algorithm for function estimation}\label{sec:boost}
In this section, we describe methods to compute step 4 in Algorithm~\ref{alg:ccf}. 
For ease of notation, we present methods for unweighted estimation, as weighted estimation does not pose technical difficulties:
\begin{equation}\label{eqn:loss}
\argmin_{\vec f \in \Omega} \sum_{i=1}^n s(y_i, f_i).
\end{equation}
Boosting techniques addressing the problem (\ref{eqn:loss}) are well-documented in the literature; see \citet{chen2016xgboost, sigrist2021gradient}, and references therein.
Briefly, the boosting solution is an additive model given by
\begin{equation}\label{eqn:add}
\hat{f}_i = F_M(\vec x_i)=\sum_{i=1}^M t_m(\vec x_i),\quad i =1, ..., n,
\end{equation}
where $F_M(\vec x_i)$ is stagewisely constructed by sequentially adding an update $t_m(\vec x_i)$ to the current estimate $F_{m-1}(\vec x_i)$:
\begin{equation}\label{eqn:boost}
F_m(\vec x_i)= F_{m-1}(\vec x_i) + t_m(\vec x_i),\quad m=1, ..., M.
\end{equation}

There are different ways to compute $\vec t_m(\vec x)=(t_m(\vec x_1), ..., t_m(\vec x_n))^\mathsf{T}$: gradient and Newton-type updates are the most popular. 
When the second derivative of the loss function exists, the Newton-type update is preferred over the gradient update to achieve fast convergence \citep{sigrist2021gradient}. 
\begin{equation}\label{eqn:tree}
\vec t_m(\vec x) = \argmin_{\vec f \in H} \sum_{i=1}^n h_{m, i}\biggl(-\frac{d_{m,i}}{h_{m,i}}-f(x_i)\biggr)^2,
\end{equation}
where the first and second derivatives of the loss function $s$ for observation $i$ are given by:
\begin{equation*}
d_{m,i}=\frac{\partial}{\partial f} s(y_i, f)|_{f=F_{m-1}(x_i)},
\end{equation*}
\begin{equation*}
h_{m,i}=\frac{\partial^2}{\partial f^2} s(y_i, f)|_{f=F_{m-1}(x_i)}.
\end{equation*}
For quadratic loss $s(y_i, f)=\frac{(y_i-f)^2}{2}$, we obtain $h_{m,i}=1$. In this case, the Newton-update is reduced to the gradient update. 
Note that $f\in\Omega$ in (\ref{eqn:loss}) and $f\in H$ in (\ref{eqn:tree}), indicating that optimal base learners can be found in     $H$ and a linear combination of them can be found in $\Omega$ through the boosting algorithm.

To prevent overfitting, boosting also implements a step-size shrinkage parameter $0 < \eta \leq 1$ in the update (\ref{eqn:boost}):
\begin{equation*}
F_m(\vec x_i)= F_{m-1}(\vec x_i) + \eta t_m(\vec x_i),\quad m=1, ..., M.
\end{equation*}
\subsection{Penalized estimation}
Another strategy to avoid overfitting is to add a regularization term to the objective function (\ref{eqn:emlos}):
\begin{equation}\label{eqn:emlos1}
    \sum_{i=1}^n \ell(y_i, {f}_i) + \sum_{m=1}^M \Lambda(t_m),
\end{equation}
where $\Lambda$ penalizes the model complexity. 

If $H$ is the linear space of linear models with a $p$-dimensional predictor, i.e., $t_m(\vec x_i) = \vec x_i ^\mathsf{T}\bm\beta_m, \bm\beta_m=(\beta_{1m}, ..., \beta_{pm})^\mathsf{T}$, let 
\begin{equation*}
\Lambda(t_m)=\frac{1}{2}\lambda\sum_{j=1}^p\beta_{jm}^2 + \alpha\sum_{j=1}^p|\beta_{jm}|,
\end{equation*}
where $\lambda \geq 0, \alpha \geq 0$ are the $L_2$ and $L_1$ regularization parameters, respectively. Note that $\Lambda(t_m)$ provides shrinkage estimators and can conduct variable selection.

Suppose that $H$ is the linear space of regression trees. Each regression tree splits the predictor space into disjoint hyper-rectangles with sides parallel to the coordinate axes \citep{wang2018robust}. 
Specifically, denote the hyper-rectangles in the $m$-th boosting iteration as $R_{jm}, j=1, ..., J$. Let $t_m(\vec x_i)=\beta_{jm}, \vec x_i \in R_{jm}, i=1, ..., n, j=1, ..., J$. With $\gamma \geq 0$, the penalty can be defined as in \citet{chen2016xgboost}:
\begin{equation*}
\Lambda(t_m)=\gamma J + \frac{1}{2}\lambda\sum_{j=1}^J\beta_{jm}^2 + \alpha\sum_{j=1}^J|\beta_{jm}|.
\end{equation*}

Consequently, the boosting estimation in Section~\ref{sec:boost} requires modifications. The     optimization problem (\ref{eqn:loss}) is changed to
\begin{equation*}\label{eqn:loss2}
\argmin_{\vec f \in \Omega} \sum_{i=1}^n s(y_i, f_i) + \sum_{m=1}^M \Lambda(t_m).
\end{equation*}
The Newton-type update (\ref{eqn:tree}) is modified accordingly:
\begin{equation*}\label{eqn:tree2}
\vec t_m(\vec x) = \argmin_{\vec f \in H} \sum_{i=1}^n h_{m, i}\left(-\frac{d_{m,i}}{h_{m,i}}-f(x_i)\right)^2 + \Lambda(t_m).
\end{equation*}
Step 4 in Algorithm~\ref{alg:ccf} can be modified accordingly.
\subsection{Implementation and tuning parameter selection}
The requirement of $z \geq 0$ on the domain of $g$, that is, $s(u) \geq 0$, may be relaxed for some concave functions $g$ in Table~\ref{tab:gs}. To satisfy all $g$ functions, however, it is simpler to require $s(u)$ to be non-negative. When $s(u) < 0$, such as in the case of a negative log-likelihood value for the gamma distribution, we can ensure a nonnegative loss by subtracting some data-dependent constant. Specifically, if $s(y, f(\vec x)) < 0$, one remedy is to subtract a constant $C$ such that $s(y, f(\vec x)) - C \geq 0$.
For the exponential family, $s(y, f(\vec{x}_i)) \geq s(y, y)$ holds, as $s(y, y)$ represents the negative log-likelihood value of a saturated model. Thus, a desired convex loss is $s(y_i, f(x_i)) - \min_{i=1, ..., n}s(y_i, y_i) \geq 0$. 

To minimize the penalized convex loss function (\ref{eqn:emlos1})
 or its weighted version in step 4 of Algorithm~\ref{alg:ccf}, we employ the popular boosting
\proglang{R} package, \pkg{xgboost}. There are two layers of iterations in the algorithm: the outer layer updates weights, and the inner layer comprises boosting iterations, where early stopping of iterations does not guarantee convergence. On the other hand, the output $\vec{f}^{(K)}$ may overfit the data. 
In practice, a two-stage process may be considered: In the first stage, Algorithm~\ref{alg:ccf} is applied to obtain optimal weights for the given observations. In the second stage, a data-driven method such as cross-validation can be used to compute a reliable estimate of errors for a weighted boosting model. This process can also be utilized to determine an optimal parameter. For instance, Algorithm~\ref{alg:cv} can be used to select an optimal robust parameter $\theta$ and the corresponding IRBoost model.
To illustrate the process, a data example
will be presented in Section~\ref{sec:reg}.
\begin{algorithm}[!htbp]
\begin{algorithmic}[1]
\caption{IRBoost Tuning Parameter Selection}\label{alg:cv}
\STATE \textbf{Input:} training samples $\{(\vec x_1, y_1), ..., (\vec x_n, y_n)\}$, 
concave component $g$ with parameters $\theta_1, ..., \theta_T$, convex component $s$, starting points $z_i, i=1, ..., n$ and iteration count $K$. 
\FOR{$t=1$ to $T$}
\STATE Compute IRBoost model $\vec{f}_t$ and robustness weights 
$-v_{it}^{(K)}, i=1, ..., n$ with $\theta_t$ using Algorithm~\ref{alg:ccf}.
\STATE Compute cross-validation errors $\epsilon_t$ using weighted boosting algorithm with weights $-v_{it}^{(K)}, i=1, ..., n$.
\ENDFOR
\STATE \textbf{Output:} $
\hat t=\argmin_{1 \leq t \leq T}\epsilon_t, \hat\theta=\theta_{\hat t}, 
\hat v_i^{(K)} = v_{i\hat t}^{(K)}, \hat{\vec{f}}=\vec{f}_{\hat t}$. 
\end{algorithmic}
\end{algorithm}
Other parameters, such as the boosting iteration $M$, penalty number $\gamma$ for trees, regularization terms $\lambda$ and $\alpha$, can be similarly chosen. 
Alternatively, since the $\theta$ parameter is typically considered a hyperparameter, a more computationally convenient approach in the literature is to conduct estimation for different values of $\theta$ and compare the results. One can begin with a large value of $\theta$ for less robust estimation and move towards smaller values of $\theta$ for more robust results.

Nonconvex loss optimization algorithms may lead to local solutions depending on the initial values. In the implementation of \pkg{irboost}, a user can specify the initial values $z_i \geq 0, i=1, ..., n$, with default values of $z_i$ being the initial weights if provided or the vector of $1$s in that order. The user may then compare the results from potentially different solutions and obtain an optimal solution afterward.

The source version of the \pkg{irboost} package is freely available from the Comprehensive \proglang{R} Archive Network (\url{http://CRAN.R-project.org}). The reader can install the package directly from the \proglang{R} prompt via:
\begin{Shaded}
\begin{Highlighting}[]
\FunctionTok{install.packages}\NormalTok{(}\StringTok{"irboost"}\NormalTok{)}
\end{Highlighting}
\end{Shaded}

All analyses presented below are contained in a package vignette. The rendered output of the analyses is available by the \proglang{R}-command:

\begin{Shaded}
\begin{Highlighting}[]
\FunctionTok{library}\NormalTok{(}\StringTok{"irboost"}\NormalTok{)}
\FunctionTok{vignette}\NormalTok{(}\StringTok{"static\_irbst"}\NormalTok{, }\AttributeTok{package =} \StringTok{"irboost"}\NormalTok{)}
\end{Highlighting}
\end{Shaded}

\hypertarget{applications}{%
\section{Applications}\label{applications}}

\hypertarget{robust-boosting-for-regression}{%
\subsection{\texorpdfstring{Robust boosting for regression\label{sec:reg}}{Robust boosting for regression}}\label{robust-boosting-for-regression}}

In this example, we predict the median value of owner-occupied homes in the suburbs of Boston, using data publicly available from the UCI machine learning data repository. The dataset comprises 506 observations and 13 predictors. \citet{wang2024unified} provides an alternative robust estimation for comparison.

\begin{Shaded}
\begin{Highlighting}[]
\NormalTok{urlname }\OtherTok{\textless{}{-}} \StringTok{"https://archive.ics.uci.edu/ml/"}
\NormalTok{filename }\OtherTok{\textless{}{-}} \StringTok{"machine{-}learning{-}databases/housing/housing.data"}
\NormalTok{dat }\OtherTok{\textless{}{-}} \FunctionTok{read.table}\NormalTok{(}\FunctionTok{paste0}\NormalTok{(urlname, filename), }\AttributeTok{sep =} \StringTok{""}\NormalTok{, }\AttributeTok{header =} \ConstantTok{FALSE}\NormalTok{)}
\NormalTok{dat }\OtherTok{\textless{}{-}} \FunctionTok{as.matrix}\NormalTok{(dat)}
\FunctionTok{colnames}\NormalTok{(dat) }\OtherTok{\textless{}{-}} \FunctionTok{c}\NormalTok{(}\StringTok{"CRIM"}\NormalTok{, }\StringTok{"ZN"}\NormalTok{, }\StringTok{"INDUS"}\NormalTok{, }\StringTok{"CHAS"}\NormalTok{, }\StringTok{"NOX"}\NormalTok{, }\StringTok{"RM"}\NormalTok{,}
    \StringTok{"AGE"}\NormalTok{, }\StringTok{"DIS"}\NormalTok{, }\StringTok{"RAD"}\NormalTok{, }\StringTok{"TAX"}\NormalTok{, }\StringTok{"PTRATIO"}\NormalTok{, }\StringTok{"B"}\NormalTok{, }\StringTok{"LSTAT"}\NormalTok{, }\StringTok{"MEDV"}\NormalTok{)}
\NormalTok{p }\OtherTok{\textless{}{-}} \FunctionTok{dim}\NormalTok{(dat)[}\DecValTok{2}\NormalTok{]}
\end{Highlighting}
\end{Shaded}

We apply IRBoost with the concave component \code{bcave} and the convex component least squares.

\begin{Shaded}
\begin{Highlighting}[]
\FunctionTok{library}\NormalTok{(}\StringTok{"irboost"}\NormalTok{)}
\NormalTok{param }\OtherTok{\textless{}{-}} \FunctionTok{list}\NormalTok{(}\AttributeTok{objective =} \StringTok{"reg:squarederror"}\NormalTok{, }\AttributeTok{max\_depth =} \DecValTok{2}\NormalTok{)}
\NormalTok{fit1 }\OtherTok{\textless{}{-}} \FunctionTok{irboost}\NormalTok{(}\AttributeTok{data =}\NormalTok{ dat[, }\SpecialCharTok{{-}}\NormalTok{p], }\AttributeTok{label =}\NormalTok{ dat[, p], }\AttributeTok{cfun =} \StringTok{"bcave"}\NormalTok{,}
    \AttributeTok{s =} \DecValTok{10}\NormalTok{, }\AttributeTok{params =}\NormalTok{ param, }\AttributeTok{verbose =} \DecValTok{0}\NormalTok{, }\AttributeTok{nrounds =} \DecValTok{50}\NormalTok{)}
\FunctionTok{plot}\NormalTok{(fit1}\SpecialCharTok{$}\NormalTok{weight\_update, }\AttributeTok{ylab =} \StringTok{"Weight"}\NormalTok{)  }\CommentTok{\# plot robustness weights}
\NormalTok{id }\OtherTok{\textless{}{-}} \FunctionTok{sort.list}\NormalTok{(fit1}\SpecialCharTok{$}\NormalTok{weight\_update)[}\DecValTok{1}\SpecialCharTok{:}\DecValTok{4}\NormalTok{]  }\CommentTok{\# 4 obs. with smallest weights}
\FunctionTok{text}\NormalTok{(id, fit1}\SpecialCharTok{$}\NormalTok{weight\_update[id] }\SpecialCharTok{{-}} \FloatTok{0.02}\NormalTok{, id, }\AttributeTok{col =} \StringTok{"red"}\NormalTok{)  }\CommentTok{\# highlight 4 obs. }
\end{Highlighting}
\end{Shaded}

\begin{figure}
\centering
\includegraphics{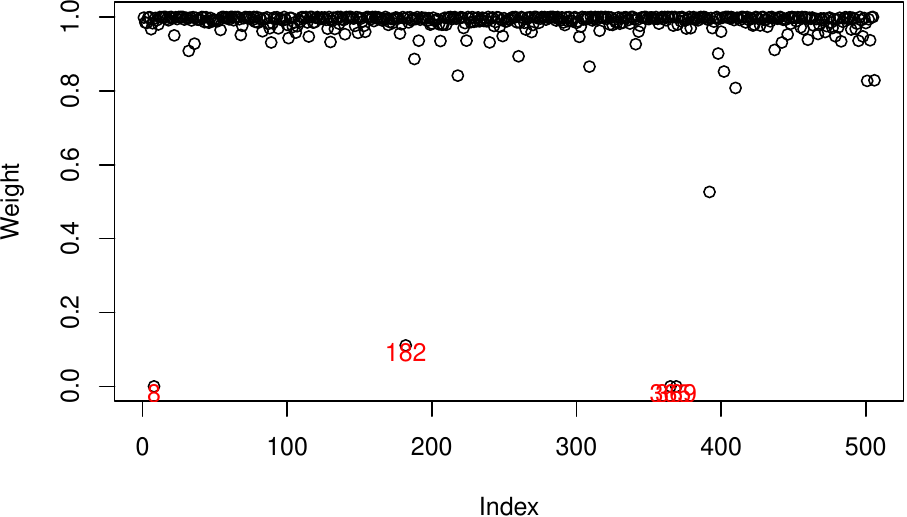}
\caption{\label{fig:weights}Robustness weights of IRBoost for the Boston housing data.}
\end{figure}

\begin{figure}

{\centering \subfloat[Observed and predicted values.\label{fig:pred-1}]{\includegraphics[width=0.5\linewidth]{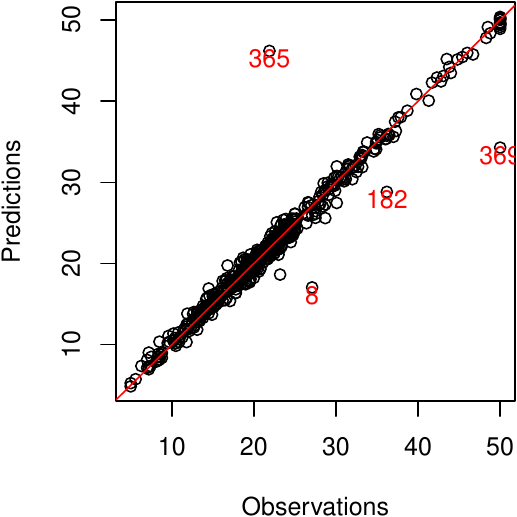} }\subfloat[Irboost and xgboost with robustness weights.\label{fig:pred-2}]{\includegraphics[width=0.5\linewidth]{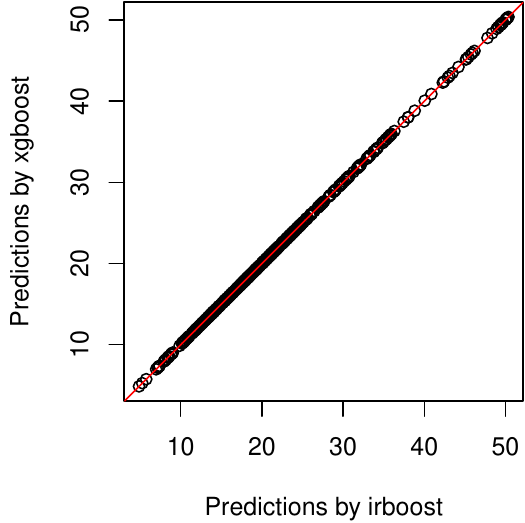} }

}

\caption{Prediction for the Boston housing data.}\label{fig:pred}
\end{figure}

Figure \ref{fig:weights} displays the observation weights used when IRBoost converges, highlighting the four smallest values, which are considered outliers. We plot the observed median housing prices against the predicted values in Figure \ref{fig:pred}(a).
Notably, the four observations with the smallest weights deviate significantly from their predicted values, but this outcome is not surprising. IRBoost returns a weighted boosting estimation. The implementation of \code{irboost} is equivalent to \code{xgboost} with weights. 
Using the weights learned from IRBoost with XGBoost results in predictions that are identical to those produced by IRBoost. This equivalence is illustrated in the following example with Figure \ref{fig:pred}(b).

\begin{Shaded}
\begin{Highlighting}[]
\FunctionTok{par}\NormalTok{(}\AttributeTok{pty =} \StringTok{"s"}\NormalTok{)}
\FunctionTok{plot}\NormalTok{(dat[, p], }\FunctionTok{predict}\NormalTok{(fit1, }\AttributeTok{newdata =}\NormalTok{ dat[, }\SpecialCharTok{{-}}\NormalTok{p]), }\AttributeTok{xlab =} \StringTok{"Observations"}\NormalTok{,}
    \AttributeTok{ylab =} \StringTok{"Predictions"}\NormalTok{)  }\CommentTok{\# obs. vs predictions }
\FunctionTok{text}\NormalTok{(dat[id, p], }\FunctionTok{predict}\NormalTok{(fit1, }\AttributeTok{newdata =}\NormalTok{ dat[id, }\SpecialCharTok{{-}}\NormalTok{p]) }\SpecialCharTok{{-}} \DecValTok{1}\NormalTok{, id,}
    \AttributeTok{col =} \StringTok{"red"}\NormalTok{)  }\CommentTok{\# highlight 4 obs. with smallest weights}
\FunctionTok{abline}\NormalTok{(}\DecValTok{0}\NormalTok{, }\DecValTok{1}\NormalTok{, }\AttributeTok{col =} \StringTok{"red"}\NormalTok{)  }\CommentTok{\# 45{-}degree line}
\FunctionTok{library}\NormalTok{(}\StringTok{"xgboost"}\NormalTok{)}
\NormalTok{fit\_xg }\OtherTok{\textless{}{-}}\NormalTok{ xgboost}\SpecialCharTok{::}\FunctionTok{xgboost}\NormalTok{(}\AttributeTok{data =}\NormalTok{ dat[, }\SpecialCharTok{{-}}\NormalTok{p], }\AttributeTok{label =}\NormalTok{ dat[, p],}
    \AttributeTok{weight =}\NormalTok{ fit1}\SpecialCharTok{$}\NormalTok{weight\_update, }\AttributeTok{params =}\NormalTok{ param, }\AttributeTok{verbose =} \DecValTok{0}\NormalTok{,}
    \AttributeTok{nrounds =}\NormalTok{ fit1}\SpecialCharTok{$}\NormalTok{niter)}
\FunctionTok{par}\NormalTok{(}\AttributeTok{pty =} \StringTok{"s"}\NormalTok{)  }\CommentTok{\# plot type square between irboost and xgboost}
\FunctionTok{plot}\NormalTok{(}\FunctionTok{predict}\NormalTok{(fit1, }\AttributeTok{newdata =}\NormalTok{ dat[, }\SpecialCharTok{{-}}\NormalTok{p]), }\FunctionTok{predict}\NormalTok{(fit\_xg, }\AttributeTok{newdata =}\NormalTok{ dat[,}
    \SpecialCharTok{{-}}\NormalTok{p]), }\AttributeTok{xlab =} \StringTok{"Predictions by irboost"}\NormalTok{, }\AttributeTok{ylab =} \StringTok{"Predictions by xgboost"}\NormalTok{)}
\FunctionTok{abline}\NormalTok{(}\DecValTok{0}\NormalTok{, }\DecValTok{1}\NormalTok{, }\AttributeTok{col =} \StringTok{"red"}\NormalTok{)  }\CommentTok{\# 45{-}degree line}
\end{Highlighting}
\end{Shaded}

We can compare computing times between \code{irboost} and \code{xgboost}. As shown below, both computing tasks completed within one second on an Intel\textsuperscript{\textregistered} Core\texttrademark{} i9-10900X CPU @ 3.70GHz \(\times\) 8 processor with 16GB of RAM. Since the former involves iterative reweighting runs of \code{xgboost}, it is expected to take more computing time than a single run of \code{xgboost}.

\begin{Shaded}
\begin{Highlighting}[]
\CommentTok{\# computing time for irboost}
\FunctionTok{system.time}\NormalTok{(}\FunctionTok{irboost}\NormalTok{(}\AttributeTok{data =}\NormalTok{ dat[, }\SpecialCharTok{{-}}\NormalTok{p], }\AttributeTok{label =}\NormalTok{ dat[, p], }\AttributeTok{cfun =} \StringTok{"bcave"}\NormalTok{,}
    \AttributeTok{s =} \DecValTok{10}\NormalTok{, }\AttributeTok{params =}\NormalTok{ param, }\AttributeTok{verbose =} \DecValTok{0}\NormalTok{, }\AttributeTok{nrounds =} \DecValTok{50}\NormalTok{))[}\StringTok{"elapsed"}\NormalTok{]}
\end{Highlighting}
\end{Shaded}

\begin{verbatim}
## elapsed 
##   0.941
\end{verbatim}

\begin{Shaded}
\begin{Highlighting}[]
\CommentTok{\# computing time for xgboost}
\FunctionTok{system.time}\NormalTok{(xgboost}\SpecialCharTok{::}\FunctionTok{xgboost}\NormalTok{(}\AttributeTok{data =}\NormalTok{ dat[, }\SpecialCharTok{{-}}\NormalTok{p], }\AttributeTok{label =}\NormalTok{ dat[,}
\NormalTok{    p], }\AttributeTok{weight =}\NormalTok{ fit1}\SpecialCharTok{$}\NormalTok{weight\_update, }\AttributeTok{params =}\NormalTok{ param, }\AttributeTok{verbose =} \DecValTok{0}\NormalTok{,}
    \AttributeTok{nrounds =}\NormalTok{ fit1}\SpecialCharTok{$}\NormalTok{niter))[}\StringTok{"elapsed"}\NormalTok{]}
\end{Highlighting}
\end{Shaded}

\begin{verbatim}
## elapsed 
##   0.127
\end{verbatim}

\begin{figure}
\centering
\includegraphics{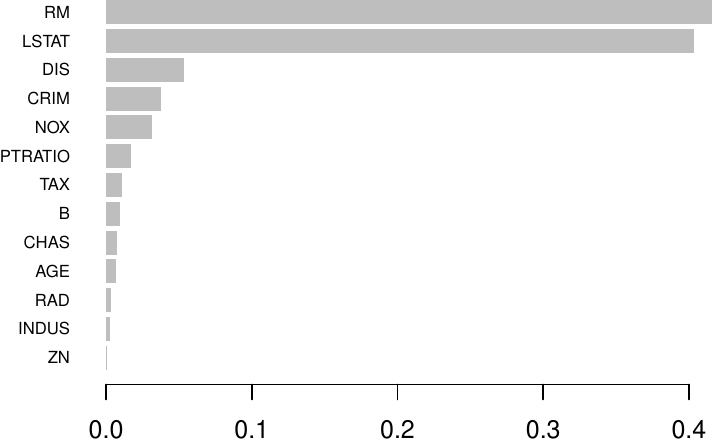}
\caption{\label{fig:imp}Variable importance measures for the Boston housing data.}
\end{figure}

Feature importance from the learned model is displayed in Figure \ref{fig:imp}. The figure reveals that the top two factors for predicting median housing prices are the average number of rooms per dwelling (RM) and the percentage values of the lower status of the population (LSTAT).

\begin{Shaded}
\begin{Highlighting}[]
\NormalTok{importance\_matrix }\OtherTok{\textless{}{-}}\NormalTok{ xgboost}\SpecialCharTok{::}\FunctionTok{xgb.importance}\NormalTok{(}\AttributeTok{model =}\NormalTok{ fit1)  }\CommentTok{\# importance metric}
\NormalTok{xgboost}\SpecialCharTok{::}\FunctionTok{xgb.plot.importance}\NormalTok{(}\AttributeTok{importance\_matrix =}\NormalTok{ importance\_matrix)  }\CommentTok{\# plot }
\end{Highlighting}
\end{Shaded}

\begin{figure}
\includegraphics[width=0.95\linewidth,height=0.3\textheight]{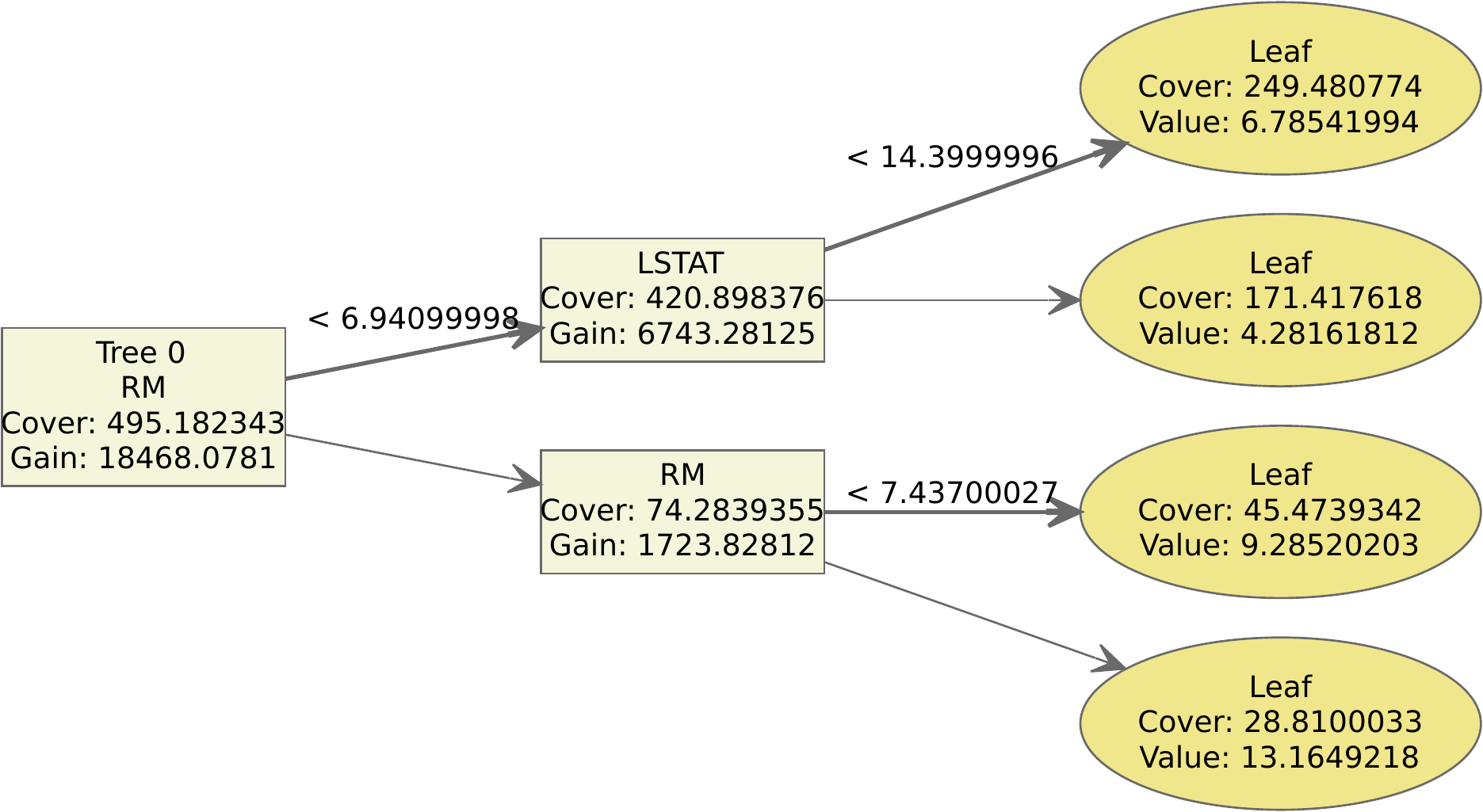} \caption{First tree in IRBoost for the Boston housing data.}\label{fig:tree}
\end{figure}

The first tree used to build the model is depicted in Figure \ref{fig:tree}.

\begin{Shaded}
\begin{Highlighting}[]
\NormalTok{xgboost}\SpecialCharTok{::}\FunctionTok{xgb.plot.tree}\NormalTok{(}\AttributeTok{model =}\NormalTok{ fit1, }\AttributeTok{trees =} \DecValTok{0}\NormalTok{)}
\end{Highlighting}
\end{Shaded}

We can optimize tuning parameters using cross-validation with the built-in functions in \pkg{xgboost}. First, update the data format with the estimated robustness weights, then determine the optimal IRBoost iteration.

\begin{Shaded}
\begin{Highlighting}[]
\NormalTok{dtrain }\OtherTok{\textless{}{-}}\NormalTok{ xgboost}\SpecialCharTok{::}\FunctionTok{xgb.DMatrix}\NormalTok{(}\AttributeTok{data =}\NormalTok{ dat[, }\SpecialCharTok{{-}}\NormalTok{p], }\AttributeTok{label =}\NormalTok{ dat[,}
\NormalTok{    p])  }\CommentTok{\# create a DMatrix for training data}
\NormalTok{xgboost}\SpecialCharTok{::}\FunctionTok{setinfo}\NormalTok{(dtrain, }\StringTok{"weight"}\NormalTok{, fit1}\SpecialCharTok{$}\NormalTok{weight\_update)  }\CommentTok{\# set weight information}
\NormalTok{param }\OtherTok{\textless{}{-}} \FunctionTok{list}\NormalTok{(}\AttributeTok{booster =} \StringTok{"gbtree"}\NormalTok{, }\AttributeTok{objective =} \StringTok{"reg:squarederror"}\NormalTok{)}
\FunctionTok{set.seed}\NormalTok{(}\DecValTok{136}\NormalTok{)  }\CommentTok{\# set the seed for reproducibility}
\NormalTok{xgbcv }\OtherTok{\textless{}{-}}\NormalTok{ xgboost}\SpecialCharTok{::}\FunctionTok{xgb.cv}\NormalTok{(}\AttributeTok{params =}\NormalTok{ param, }\AttributeTok{data =}\NormalTok{ dtrain, }\AttributeTok{nrounds =} \DecValTok{200}\NormalTok{,}
    \AttributeTok{early\_stopping\_rounds =} \DecValTok{20}\NormalTok{, }\AttributeTok{nfold =} \DecValTok{5}\NormalTok{, }\AttributeTok{prediction =} \ConstantTok{TRUE}\NormalTok{)}
\end{Highlighting}
\end{Shaded}

\begin{Shaded}
\begin{Highlighting}[]
\NormalTok{xgbcv}\SpecialCharTok{$}\NormalTok{best\_iteration}
\end{Highlighting}
\end{Shaded}

\begin{verbatim}
## [1] 37
\end{verbatim}

Continuing in the same vein, we can identify the optimal robustness parameter \(\theta\), and the corresponding IRBoost iteration. For instance, to select a preferable \(\theta\) from the set \(\{5, 10\}\), the cross-validation results below indicate that \(\theta=5\) yields a smaller root mean squared error on the test data. Moreover, this procedure identifies the optimal IRBoost iteration as 39.

\begin{Shaded}
\begin{Highlighting}[]
\NormalTok{dtrain\_cv }\OtherTok{\textless{}{-}}\NormalTok{ xgboost}\SpecialCharTok{::}\FunctionTok{xgb.DMatrix}\NormalTok{(}\AttributeTok{data =}\NormalTok{ dat[, }\SpecialCharTok{{-}}\NormalTok{p], }\AttributeTok{label =}\NormalTok{ dat[,}
\NormalTok{    p])  }\CommentTok{\# training data}
\NormalTok{robustness\_param }\OtherTok{\textless{}{-}} \FunctionTok{c}\NormalTok{(}\DecValTok{5}\NormalTok{, }\DecValTok{10}\NormalTok{)  }\CommentTok{\# two theta values}
\NormalTok{res }\OtherTok{\textless{}{-}} \ConstantTok{NULL}
\ControlFlowTok{for}\NormalTok{ (i }\ControlFlowTok{in} \DecValTok{1}\SpecialCharTok{:}\FunctionTok{length}\NormalTok{(robustness\_param)) \{}
\NormalTok{    fit\_init }\OtherTok{\textless{}{-}} \FunctionTok{irboost}\NormalTok{(}\AttributeTok{data =}\NormalTok{ dat[, }\SpecialCharTok{{-}}\NormalTok{p], }\AttributeTok{label =}\NormalTok{ dat[, p], }\AttributeTok{cfun =} \StringTok{"bcave"}\NormalTok{,}
        \AttributeTok{s =}\NormalTok{ robustness\_param[i], }\AttributeTok{params =} \FunctionTok{list}\NormalTok{(}\AttributeTok{objective =} \StringTok{"reg:squarederror"}\NormalTok{,}
            \AttributeTok{max\_depth =} \DecValTok{2}\NormalTok{), }\AttributeTok{verbose =} \DecValTok{0}\NormalTok{, }\AttributeTok{nrounds =} \DecValTok{50}\NormalTok{)  }\CommentTok{\# fit irboost model}
\NormalTok{    xgboost}\SpecialCharTok{::}\FunctionTok{setinfo}\NormalTok{(dtrain\_cv, }\StringTok{"weight"}\NormalTok{, fit\_init}\SpecialCharTok{$}\NormalTok{weight\_update)  }\CommentTok{\# new weights}
    \FunctionTok{set.seed}\NormalTok{(}\DecValTok{136}\NormalTok{)}
\NormalTok{    xgbcv }\OtherTok{\textless{}{-}}\NormalTok{ xgboost}\SpecialCharTok{::}\FunctionTok{xgb.cv}\NormalTok{(}\AttributeTok{params =}\NormalTok{ param, }\AttributeTok{data =}\NormalTok{ dtrain\_cv,}
        \AttributeTok{early\_stopping\_rounds =} \DecValTok{20}\NormalTok{, }\AttributeTok{nrounds =} \DecValTok{200}\NormalTok{, }\AttributeTok{nfold =} \DecValTok{5}\NormalTok{,}
        \AttributeTok{prediction =} \ConstantTok{TRUE}\NormalTok{)  }\CommentTok{\# 5{-}fold CV}
\NormalTok{    reslog }\OtherTok{\textless{}{-}}\NormalTok{ xgbcv}\SpecialCharTok{$}\NormalTok{evaluation\_log[xgbcv}\SpecialCharTok{$}\NormalTok{best\_iteration]  }\CommentTok{\# best values in CV}
\NormalTok{    tmp }\OtherTok{\textless{}{-}} \FunctionTok{unlist}\NormalTok{(}\FunctionTok{c}\NormalTok{(}\AttributeTok{theta =}\NormalTok{ robustness\_param[i], reslog))  }\CommentTok{\# combine theta}
\NormalTok{    res }\OtherTok{\textless{}{-}} \FunctionTok{rbind}\NormalTok{(res, tmp)  }\CommentTok{\# combine results from previous theta}
    \FunctionTok{rownames}\NormalTok{(res) }\OtherTok{\textless{}{-}} \ConstantTok{NULL}
\NormalTok{\}}
\end{Highlighting}
\end{Shaded}

\begin{Shaded}
\begin{Highlighting}[]
\FunctionTok{print}\NormalTok{(res, }\AttributeTok{digits =} \DecValTok{2}\NormalTok{)}
\end{Highlighting}
\end{Shaded}

\begin{verbatim}
##      theta iter train_rmse_mean train_rmse_std test_rmse_mean
## [1,]     5   39            0.30          0.019            2.3
## [2,]    10   37            0.37          0.036            3.0
##      test_rmse_std
## [1,]          0.31
## [2,]          0.33
\end{verbatim}

\hypertarget{robust-logistic-boosting}{%
\subsection{Robust logistic boosting}\label{robust-logistic-boosting}}

A binary classification problem, as proposed by \citet{long2010random}, involves a response variable \(y\) randomly chosen to be -1 or +1 with equal probability. Symbols A, B, and C are randomly generated with probabilities 0.25, 0.25, and 0.5, respectively. The predictor vector \(\vec x\) with 21 elements is generated as follows: if A is obtained, \(x_j=y\) for \(j=1, ..., 21\). If B is generated, \(x_j=y\) for \(j=1, ..., 11\), and \(x_j=-y\) for \(j=12, ..., 21\). If C is generated, \(x_j=y\), where \(j\)
is randomly chosen from the range of 1 to 11 with a selection of 5 elements, and from the range of 12 to 21 with a selection of 6 elements. For the remaining \(j \in \{1, 2, 3, ..., 21\}\), \(x_j=-y\). The training data is generated with \(n=400\) samples, and the test data with \(n=200\) samples.

We fit a robust logistic boosting model with the concave component \code{acave}, setting the maximum depth of a tree to 5. With a large parameter \(\theta=100\), the robustness weights are very close to \(1\) as shown below.

\begin{Shaded}
\begin{Highlighting}[]
\FunctionTok{set.seed}\NormalTok{(}\DecValTok{1947}\NormalTok{)}
\NormalTok{dat2 }\OtherTok{\textless{}{-}} \FunctionTok{dataLS}\NormalTok{(}\AttributeTok{ntr =} \DecValTok{400}\NormalTok{, }\AttributeTok{nte =} \DecValTok{200}\NormalTok{, }\AttributeTok{percon =} \DecValTok{0}\NormalTok{)  }\CommentTok{\# percon=0 means clean data}
\NormalTok{param }\OtherTok{\textless{}{-}} \FunctionTok{list}\NormalTok{(}\AttributeTok{objective =} \StringTok{"binary:logitraw"}\NormalTok{, }\AttributeTok{max\_depth =} \DecValTok{5}\NormalTok{)}
\NormalTok{fit2 }\OtherTok{\textless{}{-}} \FunctionTok{irboost}\NormalTok{(}\AttributeTok{data =}\NormalTok{ dat2}\SpecialCharTok{$}\NormalTok{xtr, }\AttributeTok{label =}\NormalTok{ dat2}\SpecialCharTok{$}\NormalTok{ytr, }\AttributeTok{cfun =} \StringTok{"acave"}\NormalTok{,}
    \AttributeTok{s =} \DecValTok{100}\NormalTok{, }\AttributeTok{params =}\NormalTok{ param, }\AttributeTok{verbose =} \DecValTok{0}\NormalTok{, }\AttributeTok{nrounds =} \DecValTok{100}\NormalTok{)}
\FunctionTok{range}\NormalTok{(fit2}\SpecialCharTok{$}\NormalTok{weight\_update)  }\CommentTok{\# range of robustness weights}
\end{Highlighting}
\end{Shaded}

\begin{verbatim}
## [1] 0.9999975 1.0000000
\end{verbatim}

\begin{figure}
\centering
\includegraphics{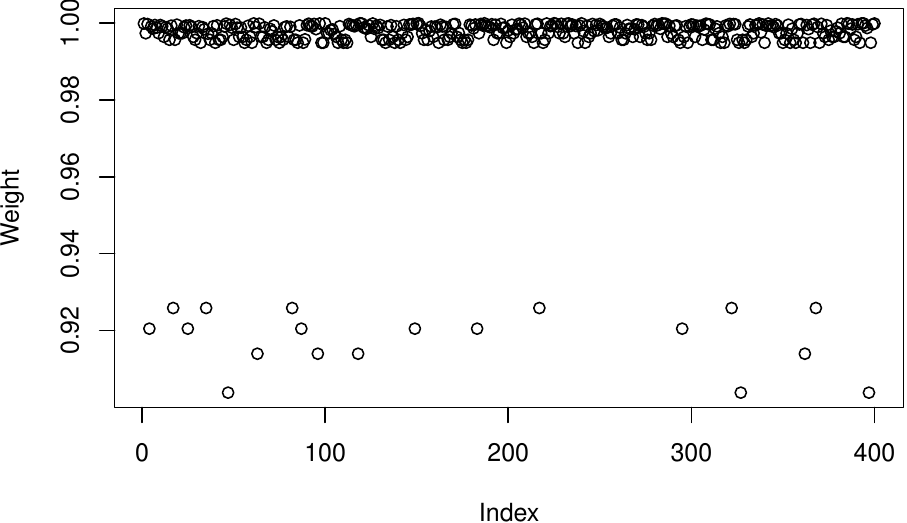}
\caption{\label{fig:weight3}Robustness weights of IRBoost with \(\theta=3\) for the contaminated simulation data.}
\end{figure}

To add outliers, we simulate data with 10\% contamination in the response variables of the training data and then apply IRBoost. Figure \ref{fig:weight3} displays the robustness weights obtained from the algorithm.

\begin{Shaded}
\begin{Highlighting}[]
\FunctionTok{set.seed}\NormalTok{(}\DecValTok{158}\NormalTok{)}
\NormalTok{dat3 }\OtherTok{\textless{}{-}} \FunctionTok{dataLS}\NormalTok{(}\AttributeTok{ntr =} \DecValTok{400}\NormalTok{, }\AttributeTok{nte =} \DecValTok{200}\NormalTok{, }\AttributeTok{percon =} \FloatTok{0.1}\NormalTok{)  }\CommentTok{\# 10\% data contamination}
\NormalTok{param }\OtherTok{\textless{}{-}} \FunctionTok{list}\NormalTok{(}\AttributeTok{objective =} \StringTok{"binary:logitraw"}\NormalTok{, }\AttributeTok{max\_depth =} \DecValTok{5}\NormalTok{)}
\NormalTok{fit3 }\OtherTok{\textless{}{-}} \FunctionTok{irboost}\NormalTok{(}\AttributeTok{data =}\NormalTok{ dat3}\SpecialCharTok{$}\NormalTok{xtr, }\AttributeTok{label =}\NormalTok{ dat3}\SpecialCharTok{$}\NormalTok{ytr, }\AttributeTok{cfun =} \StringTok{"acave"}\NormalTok{,}
    \AttributeTok{s =} \DecValTok{3}\NormalTok{, }\AttributeTok{params =}\NormalTok{ param, }\AttributeTok{verbose =} \DecValTok{0}\NormalTok{, }\AttributeTok{nrounds =} \DecValTok{100}\NormalTok{)}
\FunctionTok{plot}\NormalTok{(fit3}\SpecialCharTok{$}\NormalTok{weight\_update, }\AttributeTok{ylab =} \StringTok{"Weight"}\NormalTok{)  }\CommentTok{\# plot robustness weights}
\end{Highlighting}
\end{Shaded}

\begin{figure}
\centering
\includegraphics{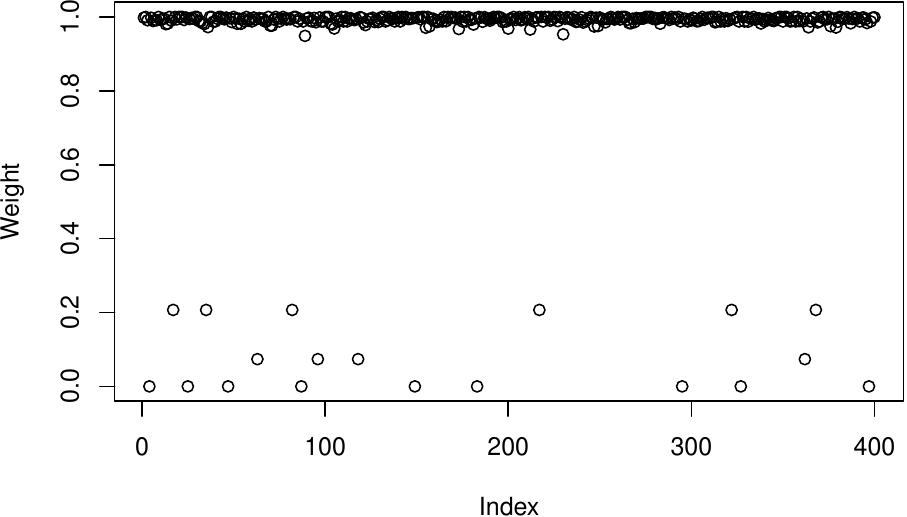}
\caption{\label{fig:weight4}Robustness weights of IRBoost with \(\theta=1\) for the contaminated simulation data.}
\end{figure}

In the third robust logistic boosting, we set the robustness hyperparameter value \(\theta\) to 1 (\code{s=1} in the \code{irboost} function) for a more robust estimation. Consequently, certain observations exhibit decreased weights, as illustrated in Figure \ref{fig:weight4}.

\begin{Shaded}
\begin{Highlighting}[]
\NormalTok{param }\OtherTok{\textless{}{-}} \FunctionTok{list}\NormalTok{(}\AttributeTok{objective =} \StringTok{"binary:logitraw"}\NormalTok{, }\AttributeTok{max\_depth =} \DecValTok{5}\NormalTok{)}
\NormalTok{fit4 }\OtherTok{\textless{}{-}} \FunctionTok{irboost}\NormalTok{(}\AttributeTok{data =}\NormalTok{ dat3}\SpecialCharTok{$}\NormalTok{xtr, }\AttributeTok{label =}\NormalTok{ dat3}\SpecialCharTok{$}\NormalTok{ytr, }\AttributeTok{cfun =} \StringTok{"acave"}\NormalTok{,}
    \AttributeTok{s =} \DecValTok{1}\NormalTok{, }\AttributeTok{params =}\NormalTok{ param, }\AttributeTok{verbose =} \DecValTok{0}\NormalTok{, }\AttributeTok{nrounds =} \DecValTok{100}\NormalTok{)}
\FunctionTok{plot}\NormalTok{(fit4}\SpecialCharTok{$}\NormalTok{weight\_update, }\AttributeTok{ylab =} \StringTok{"Weight"}\NormalTok{)  }\CommentTok{\# plot robustness weights}
\end{Highlighting}
\end{Shaded}

\begin{figure}
\centering
\includegraphics{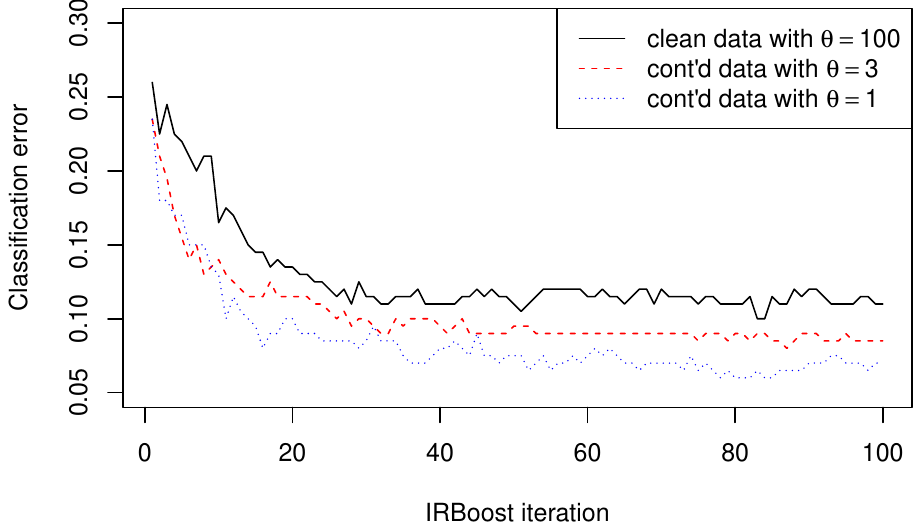}
\caption{\label{fig:error4}Classification errors and IRBoost iterations for the simulation data.}
\end{figure}

The prediction accuracy can be compared for different models. The prediction error of the test data at each IRBoost iteration is depicted in Figure \ref{fig:error4}, demonstrating that the most accurate predictions come from the robust model, even in the presence of outliers.
\begin{Shaded}
\begin{Highlighting}[]
\NormalTok{err2 }\OtherTok{\textless{}{-}}\NormalTok{ err3 }\OtherTok{\textless{}{-}}\NormalTok{ err4 }\OtherTok{\textless{}{-}} \FunctionTok{rep}\NormalTok{(}\ConstantTok{NA}\NormalTok{, }\DecValTok{100}\NormalTok{)}
\ControlFlowTok{for}\NormalTok{ (i }\ControlFlowTok{in} \DecValTok{1}\SpecialCharTok{:}\DecValTok{100}\NormalTok{) \{}
\NormalTok{    pred2 }\OtherTok{\textless{}{-}} \FunctionTok{predict}\NormalTok{(fit2, }\AttributeTok{newdata =}\NormalTok{ dat2}\SpecialCharTok{$}\NormalTok{xte, }\AttributeTok{iterationrange =} \FunctionTok{c}\NormalTok{(}\DecValTok{1}\NormalTok{,}
\NormalTok{        i }\SpecialCharTok{+} \DecValTok{1}\NormalTok{))  }\CommentTok{\# prediction with the first i trees}
\NormalTok{    err2[i] }\OtherTok{\textless{}{-}} \FunctionTok{mean}\NormalTok{(}\FunctionTok{sign}\NormalTok{(pred2) }\SpecialCharTok{!=}\NormalTok{ dat2}\SpecialCharTok{$}\NormalTok{yte)  }\CommentTok{\# error at iteration i}
\NormalTok{    pred3 }\OtherTok{\textless{}{-}} \FunctionTok{predict}\NormalTok{(fit3, }\AttributeTok{newdata =}\NormalTok{ dat3}\SpecialCharTok{$}\NormalTok{xte, }\AttributeTok{iterationrange =} \FunctionTok{c}\NormalTok{(}\DecValTok{1}\NormalTok{,}
\NormalTok{        i }\SpecialCharTok{+} \DecValTok{1}\NormalTok{))}
\NormalTok{    err3[i] }\OtherTok{\textless{}{-}} \FunctionTok{mean}\NormalTok{(}\FunctionTok{sign}\NormalTok{(pred3) }\SpecialCharTok{!=}\NormalTok{ dat3}\SpecialCharTok{$}\NormalTok{yte)}
\NormalTok{    pred4 }\OtherTok{\textless{}{-}} \FunctionTok{predict}\NormalTok{(fit4, }\AttributeTok{newdata =}\NormalTok{ dat3}\SpecialCharTok{$}\NormalTok{xte, }\AttributeTok{iterationrange =} \FunctionTok{c}\NormalTok{(}\DecValTok{1}\NormalTok{,}
\NormalTok{        i }\SpecialCharTok{+} \DecValTok{1}\NormalTok{))}
\NormalTok{    err4[i] }\OtherTok{\textless{}{-}} \FunctionTok{mean}\NormalTok{(}\FunctionTok{sign}\NormalTok{(pred4) }\SpecialCharTok{!=}\NormalTok{ dat3}\SpecialCharTok{$}\NormalTok{yte)}
\NormalTok{\}}
\FunctionTok{plot}\NormalTok{(err2[}\DecValTok{1}\SpecialCharTok{:}\DecValTok{100}\NormalTok{], }\AttributeTok{ylim =} \FunctionTok{c}\NormalTok{(}\FloatTok{0.05}\NormalTok{, }\FloatTok{0.3}\NormalTok{), }\AttributeTok{type =} \StringTok{"l"}\NormalTok{, }\AttributeTok{xlab =} \StringTok{"IRBoost iteration"}\NormalTok{,}
    \AttributeTok{ylab =} \StringTok{"Classification error"}\NormalTok{)}
\FunctionTok{points}\NormalTok{(err3[}\DecValTok{1}\SpecialCharTok{:}\DecValTok{100}\NormalTok{], }\AttributeTok{col =} \StringTok{"red"}\NormalTok{, }\AttributeTok{type =} \StringTok{"l"}\NormalTok{, }\AttributeTok{lty =} \StringTok{"dashed"}\NormalTok{)}
\FunctionTok{points}\NormalTok{(err4[}\DecValTok{1}\SpecialCharTok{:}\DecValTok{100}\NormalTok{], }\AttributeTok{col =} \StringTok{"blue"}\NormalTok{, }\AttributeTok{type =} \StringTok{"l"}\NormalTok{, }\AttributeTok{lty =} \StringTok{"dotted"}\NormalTok{)}
\FunctionTok{legend}\NormalTok{(}\StringTok{"topright"}\NormalTok{, }\AttributeTok{lty =} \FunctionTok{c}\NormalTok{(}\StringTok{"solid"}\NormalTok{, }\StringTok{"dashed"}\NormalTok{, }\StringTok{"dotted"}\NormalTok{), }\AttributeTok{col =} \FunctionTok{c}\NormalTok{(}\StringTok{"black"}\NormalTok{,}
    \StringTok{"red"}\NormalTok{, }\StringTok{"blue"}\NormalTok{), }\AttributeTok{legend =} \FunctionTok{c}\NormalTok{(}\FunctionTok{expression}\NormalTok{(}\FunctionTok{paste}\NormalTok{(}\StringTok{"clean data with "}\NormalTok{,}
\NormalTok{    theta }\SpecialCharTok{==} \DecValTok{100}\NormalTok{)), }\FunctionTok{expression}\NormalTok{(}\FunctionTok{paste}\NormalTok{(}\StringTok{"cont\textquotesingle{}d data with "}\NormalTok{, theta }\SpecialCharTok{==}
    \DecValTok{3}\NormalTok{)), }\FunctionTok{expression}\NormalTok{(}\FunctionTok{paste}\NormalTok{(}\StringTok{"cont\textquotesingle{}d data with "}\NormalTok{, theta }\SpecialCharTok{==} \DecValTok{1}\NormalTok{))))}
\end{Highlighting}
\end{Shaded}

\hypertarget{robust-multiclass-boosting}{%
\subsection{Robust multiclass boosting}\label{robust-multiclass-boosting}}

\begin{figure}
\centering
\includegraphics{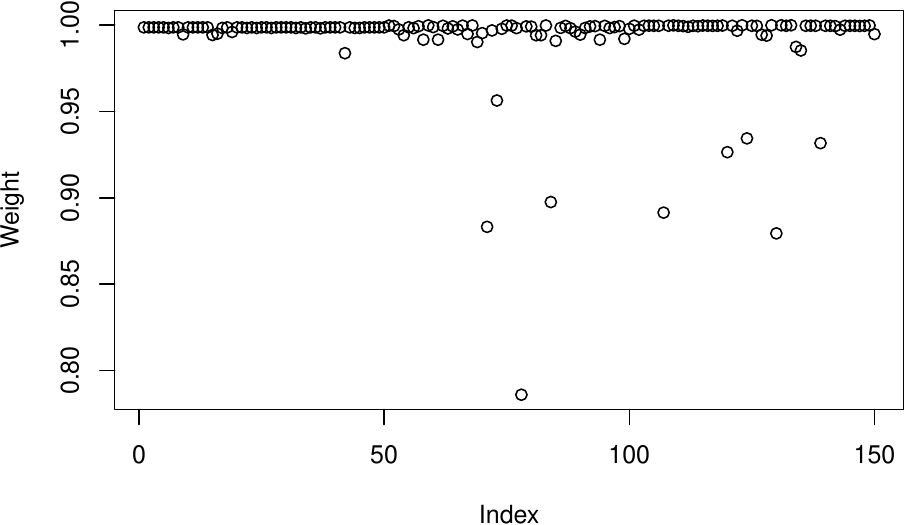}
\caption{\label{fig:weight5}Robustness weights of IRBoost for the iris data.}
\end{figure}

In a 3-class classification using the \code{iris} dataset, we run IRBoost with concave function \code{acave} and \(\theta=1\).
The robustness weights are illustrated in Figure \ref{fig:weight5}, and the model achieves perfect prediction.

\begin{Shaded}
\begin{Highlighting}[]
\NormalTok{lb }\OtherTok{\textless{}{-}} \FunctionTok{as.numeric}\NormalTok{(iris}\SpecialCharTok{$}\NormalTok{Species) }\SpecialCharTok{{-}} \DecValTok{1}  \CommentTok{\# convert text to numeric values}
\NormalTok{num\_class }\OtherTok{\textless{}{-}} \DecValTok{3}
\FunctionTok{set.seed}\NormalTok{(}\DecValTok{11}\NormalTok{)}
\NormalTok{param }\OtherTok{\textless{}{-}} \FunctionTok{list}\NormalTok{(}\AttributeTok{objective =} \StringTok{"multi:softprob"}\NormalTok{, }\AttributeTok{max\_depth =} \DecValTok{4}\NormalTok{, }\AttributeTok{eta =} \FloatTok{0.5}\NormalTok{,}
    \AttributeTok{nthread =} \DecValTok{2}\NormalTok{, }\AttributeTok{subsample =} \FloatTok{0.5}\NormalTok{, }\AttributeTok{num\_class =}\NormalTok{ num\_class)}
\NormalTok{fit5 }\OtherTok{\textless{}{-}} \FunctionTok{irboost}\NormalTok{(}\AttributeTok{data =} \FunctionTok{as.matrix}\NormalTok{(iris[, }\SpecialCharTok{{-}}\DecValTok{5}\NormalTok{]), }\AttributeTok{label =}\NormalTok{ lb, }\AttributeTok{cfun =} \StringTok{"acave"}\NormalTok{,}
    \AttributeTok{s =} \DecValTok{1}\NormalTok{, }\AttributeTok{params =}\NormalTok{ param, }\AttributeTok{verbose =} \DecValTok{0}\NormalTok{, }\AttributeTok{nrounds =} \DecValTok{10}\NormalTok{)}
\end{Highlighting}
\end{Shaded}

\begin{Shaded}
\begin{Highlighting}[]
\FunctionTok{plot}\NormalTok{(fit5}\SpecialCharTok{$}\NormalTok{weight\_update, }\AttributeTok{ylab =} \StringTok{"Weight"}\NormalTok{)  }\CommentTok{\# plot robustness weights}
\end{Highlighting}
\end{Shaded}

\begin{Shaded}
\begin{Highlighting}[]
\CommentTok{\# compute num\_class probabilities per case}
\NormalTok{pred5 }\OtherTok{\textless{}{-}} \FunctionTok{predict}\NormalTok{(fit5, }\AttributeTok{newdata =} \FunctionTok{as.matrix}\NormalTok{(iris[, }\SpecialCharTok{{-}}\DecValTok{5}\NormalTok{]))}
\CommentTok{\# reshape to a num\_class{-}columns matrix}
\NormalTok{pred5 }\OtherTok{\textless{}{-}} \FunctionTok{matrix}\NormalTok{(pred5, }\AttributeTok{ncol =}\NormalTok{ num\_class, }\AttributeTok{byrow =} \ConstantTok{TRUE}\NormalTok{)}
\NormalTok{pred5\_labels }\OtherTok{\textless{}{-}} \FunctionTok{max.col}\NormalTok{(pred5) }\SpecialCharTok{{-}} \DecValTok{1}  \CommentTok{\# probabilities to labels}
\FunctionTok{sum}\NormalTok{(pred5\_labels }\SpecialCharTok{!=}\NormalTok{ lb)  }\CommentTok{\# classification errors}
\end{Highlighting}
\end{Shaded}

\begin{verbatim}
## [1] 0
\end{verbatim}

\hypertarget{robust-poisson-boosting}{%
\subsection{Robust Poisson boosting}\label{robust-poisson-boosting}}

\begin{figure}
\centering
\includegraphics{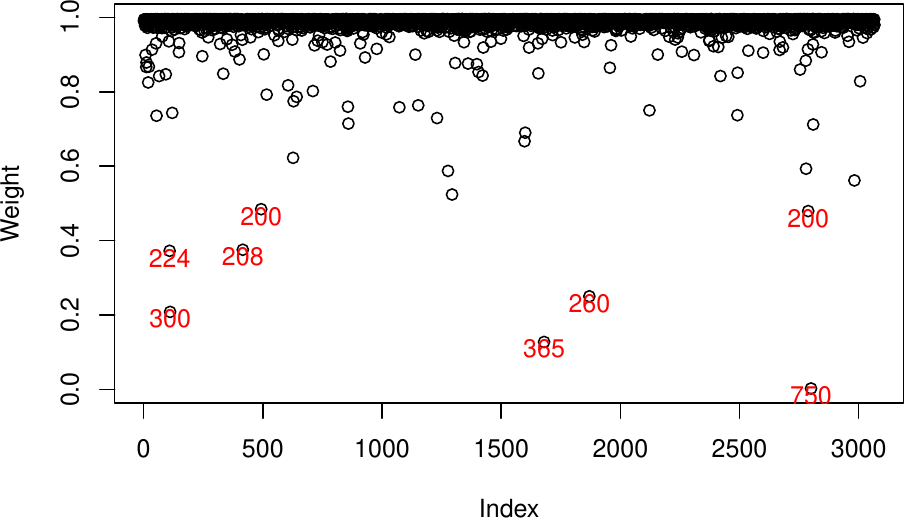}
\caption{\label{fig:weight6}Robustness weights of IRBoost for the doctor office visits data.}
\end{figure}

A survey, collected from 3066 Americans, studied health care utilization \citep{heritier2009robust, wang2024unified}. The dataset includes information on doctor office visits and 24 risk factors. A robust Poisson boosting model is fitted with the concave component \code{ccave}, and the estimated robustness weights are illustrated in Figure \ref{fig:weight6}. Doctor office visits ranging from 200 to 750 in two years are highlighted for the 8 patients with the smallest weights.

\begin{Shaded}
\begin{Highlighting}[]
\FunctionTok{data}\NormalTok{(docvisits, }\AttributeTok{package =} \StringTok{"mpath"}\NormalTok{)}
\CommentTok{\# convert factors and other types of variables into a}
\CommentTok{\# numeric matrix}
\NormalTok{x }\OtherTok{\textless{}{-}} \FunctionTok{model.matrix}\NormalTok{(}\SpecialCharTok{\textasciitilde{}}\NormalTok{age }\SpecialCharTok{+} \FunctionTok{factor}\NormalTok{(gender) }\SpecialCharTok{+} \FunctionTok{factor}\NormalTok{(race) }\SpecialCharTok{+} \FunctionTok{factor}\NormalTok{(hispan) }\SpecialCharTok{+}
    \FunctionTok{factor}\NormalTok{(marital) }\SpecialCharTok{+} \FunctionTok{factor}\NormalTok{(arthri) }\SpecialCharTok{+} \FunctionTok{factor}\NormalTok{(cancer) }\SpecialCharTok{+} \FunctionTok{factor}\NormalTok{(hipress) }\SpecialCharTok{+}
    \FunctionTok{factor}\NormalTok{(diabet) }\SpecialCharTok{+} \FunctionTok{factor}\NormalTok{(lung) }\SpecialCharTok{+} \FunctionTok{factor}\NormalTok{(hearth) }\SpecialCharTok{+} \FunctionTok{factor}\NormalTok{(stroke) }\SpecialCharTok{+}
    \FunctionTok{factor}\NormalTok{(psych) }\SpecialCharTok{+} \FunctionTok{factor}\NormalTok{(iadla) }\SpecialCharTok{+} \FunctionTok{factor}\NormalTok{(adlwa) }\SpecialCharTok{+}\NormalTok{ edyears }\SpecialCharTok{+}
\NormalTok{    feduc }\SpecialCharTok{+}\NormalTok{ meduc }\SpecialCharTok{+} \FunctionTok{log}\NormalTok{(income }\SpecialCharTok{+} \DecValTok{1}\NormalTok{) }\SpecialCharTok{+} \FunctionTok{factor}\NormalTok{(insur) }\SpecialCharTok{+} \DecValTok{0}\NormalTok{, }\AttributeTok{data =}\NormalTok{ docvisits)}
\NormalTok{param }\OtherTok{\textless{}{-}} \FunctionTok{list}\NormalTok{(}\AttributeTok{objective =} \StringTok{"count:poisson"}\NormalTok{, }\AttributeTok{max\_depth =} \DecValTok{1}\NormalTok{)}
\NormalTok{fit6 }\OtherTok{\textless{}{-}} \FunctionTok{irboost}\NormalTok{(}\AttributeTok{data =}\NormalTok{ x, }\AttributeTok{label =}\NormalTok{ docvisits}\SpecialCharTok{$}\NormalTok{visits, }\AttributeTok{cfun =} \StringTok{"ccave"}\NormalTok{,}
    \AttributeTok{s =} \DecValTok{20}\NormalTok{, }\AttributeTok{params =}\NormalTok{ param, }\AttributeTok{verbose =} \DecValTok{0}\NormalTok{, }\AttributeTok{nrounds =} \DecValTok{50}\NormalTok{)}
\end{Highlighting}
\end{Shaded}

\begin{Shaded}
\begin{Highlighting}[]
\FunctionTok{plot}\NormalTok{(fit6}\SpecialCharTok{$}\NormalTok{weight\_update, }\AttributeTok{ylab =} \StringTok{"Weight"}\NormalTok{)  }\CommentTok{\# plot robustness weights}
\NormalTok{id }\OtherTok{\textless{}{-}} \FunctionTok{sort.list}\NormalTok{(fit6}\SpecialCharTok{$}\NormalTok{weight\_update)[}\DecValTok{1}\SpecialCharTok{:}\DecValTok{8}\NormalTok{]  }\CommentTok{\# 8 obs. with smallest weights}
\FunctionTok{text}\NormalTok{(id, fit6}\SpecialCharTok{$}\NormalTok{weight\_update[id] }\SpecialCharTok{{-}} \FloatTok{0.02}\NormalTok{, docvisits}\SpecialCharTok{$}\NormalTok{visits[id],}
    \AttributeTok{col =} \StringTok{"red"}\NormalTok{)  }\CommentTok{\# highlight 8 obs.}
\end{Highlighting}
\end{Shaded}

\hypertarget{robust-survival-boosting-with-accelerated-failure-time-model}{%
\subsection{Robust survival boosting with accelerated failure time model}\label{robust-survival-boosting-with-accelerated-failure-time-model}}

\begin{figure}
\centering
\includegraphics{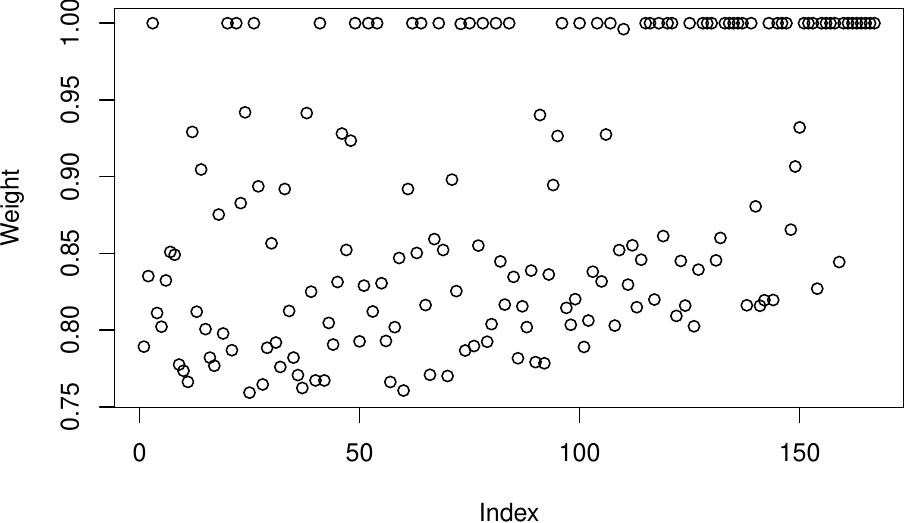}
\caption{\label{fig:weight7}Robustness weights of IRBoost for the lung cancer data.}
\end{figure}

Cox regression in survival analysis is based on a partial likelihood function. A robust extension in the CC-family is beyond the scope of this article. Alternatively, one may apply robust survival regression with the accelerated failure time model in \pkg{irboost}. The following code provides robust survival analysis for patients with advanced lung cancer from the North Central Cancer Treatment Group. Model performance is evaluated with multiple measures for survival data. Figure \ref{fig:weight7} illustrates the robustness weights using IRBoost.

\begin{Shaded}
\begin{Highlighting}[]
\FunctionTok{library}\NormalTok{(}\StringTok{"survival"}\NormalTok{)}
\NormalTok{lung1 }\OtherTok{\textless{}{-}}\NormalTok{ lung[}\FunctionTok{complete.cases}\NormalTok{(lung), ]  }\CommentTok{\# remove missing data}
\NormalTok{y\_upper\_bound }\OtherTok{\textless{}{-}} \FunctionTok{rep}\NormalTok{(}\ConstantTok{NA}\NormalTok{, }\FunctionTok{dim}\NormalTok{(lung1)[}\DecValTok{1}\NormalTok{])}
\CommentTok{\# set right{-}censoring obs. to y\_upper\_bound = Inf}
\ControlFlowTok{for}\NormalTok{ (i }\ControlFlowTok{in} \DecValTok{1}\SpecialCharTok{:}\FunctionTok{dim}\NormalTok{(lung1)[}\DecValTok{1}\NormalTok{]) \{}
    \ControlFlowTok{if}\NormalTok{ (lung1}\SpecialCharTok{$}\NormalTok{status[i] }\SpecialCharTok{==} \DecValTok{2}\NormalTok{) \{}
\NormalTok{        y\_upper\_bound[i] }\OtherTok{\textless{}{-}}\NormalTok{ lung1}\SpecialCharTok{$}\NormalTok{time[i]}
\NormalTok{    \} }\ControlFlowTok{else}\NormalTok{ y\_upper\_bound[i] }\OtherTok{\textless{}{-}} \ConstantTok{Inf}
\NormalTok{\}}
\NormalTok{x }\OtherTok{\textless{}{-}} \FunctionTok{as.matrix}\NormalTok{(lung1[, }\SpecialCharTok{!}\FunctionTok{names}\NormalTok{(lung1) }\SpecialCharTok{\%in\%} \FunctionTok{c}\NormalTok{(}\StringTok{"time"}\NormalTok{, }\StringTok{"status"}\NormalTok{)])  }\CommentTok{\# predictors}
\NormalTok{dtrain }\OtherTok{\textless{}{-}}\NormalTok{ xgboost}\SpecialCharTok{::}\FunctionTok{xgb.DMatrix}\NormalTok{(}\AttributeTok{data =}\NormalTok{ x, }\AttributeTok{label\_lower\_bound =}\NormalTok{ lung1}\SpecialCharTok{$}\NormalTok{time,}
    \AttributeTok{label\_upper\_bound =}\NormalTok{ y\_upper\_bound)  }\CommentTok{\# input data format}
\NormalTok{param }\OtherTok{\textless{}{-}} \FunctionTok{list}\NormalTok{(}\AttributeTok{objective =} \StringTok{"survival:aft"}\NormalTok{, }\AttributeTok{eval\_metric =} \StringTok{"aft{-}nloglik"}\NormalTok{,}
    \AttributeTok{aft\_loss\_distribution =} \StringTok{"normal"}\NormalTok{, }\AttributeTok{aft\_loss\_distribution\_scale =} \FloatTok{1.2}\NormalTok{,}
    \AttributeTok{max\_depth =} \DecValTok{3}\NormalTok{)}
\FunctionTok{library}\NormalTok{(}\StringTok{"Hmisc"}\NormalTok{)}
\NormalTok{fit7 }\OtherTok{\textless{}{-}} \FunctionTok{irb.train}\NormalTok{(}\AttributeTok{params =}\NormalTok{ param, }\AttributeTok{data =}\NormalTok{ dtrain, }\AttributeTok{cfun =} \StringTok{"hcave"}\NormalTok{,}
    \AttributeTok{s =} \DecValTok{3}\NormalTok{, }\AttributeTok{nrounds =} \DecValTok{50}\NormalTok{)}
\CommentTok{\# evaluate model prediction accuracy}
\NormalTok{Hmisc}\SpecialCharTok{::}\FunctionTok{rcorr.cens}\NormalTok{(}\FunctionTok{predict}\NormalTok{(fit7, }\AttributeTok{newdata =}\NormalTok{ dtrain), }\FunctionTok{Surv}\NormalTok{(}\AttributeTok{time =}\NormalTok{ lung1}\SpecialCharTok{$}\NormalTok{time,}
    \AttributeTok{event =}\NormalTok{ lung1}\SpecialCharTok{$}\NormalTok{status))}
\end{Highlighting}
\end{Shaded}

\begin{verbatim}
##        C Index            Dxy           S.D.              n 
##   9.805945e-01   9.611889e-01   5.940054e-03   1.670000e+02 
##        missing     uncensored Relevant Pairs     Concordant 
##   0.000000e+00   1.200000e+02   2.112800e+04   2.071800e+04 
##      Uncertain 
##   6.572000e+03
\end{verbatim}

\begin{Shaded}
\begin{Highlighting}[]
\FunctionTok{plot}\NormalTok{(fit7}\SpecialCharTok{$}\NormalTok{weight\_update, }\AttributeTok{ylab =} \StringTok{"Weight"}\NormalTok{)  }\CommentTok{\# plot robustness weights}
\end{Highlighting}
\end{Shaded}

 % to include knitr code from irbst_knitr.Rmd
\section{Discussion}
In this article, we introduced IRBoost as a unified robust boosting algorithm and demonstrated its versatility in regression, generalized linear models, classification, and time-to-event data analysis. The method served well for outlier detection and yielded more robust predictive models. 
Leveraging existing weighted boosting software, the approach could conveniently conduct tuning parameter selections and further explore the developed models for variable importance. The \proglang{R} package \pkg{irboost} proved to be a valuable tool in machine learning applications.

To make the IRBoost algorithm scalable for large-scale data sets, one direction is to use the stochastic majorization-minimization schemes \citep{mairal2013stochastic}.
These stochastic variants of the MM algorithm can efficiently optimize the objective function by using only a subset of the data at each iteration, rather than the entire dataset. This allows the algorithm to scale to very large problem sizes. Stochastic IRBoost remains a future research topic. The current \proglang{R} package \pkg{irboost} does not support user extensibility with other distributions. However, such functionality may be developed in a future release of the package.
\begin{acknowledgement}[title={Acknowledgments}]
The author would like to thank the Editor and an Associate Editor for their constructive comments, which substantially help improve an early draft of the manuscript.
This work was partially supported by the National Institute of Diabetes and Digestive and Kidney         Diseases of the National Institutes of Health under Award Number R21DK130006. The author expresses gratitude for the contributions of the \proglang{R} package \pkg{xgboost} to this work.
\end{acknowledgement}

%\section*{Supplementary Material}
%The R code necessary to reproduce the analysis presented in the manuscript is provided.
\bibliographystyle{jds}
\bibliography{irbst}

\end{document}